\documentclass{article}

\input{tcilatex}
\begin{document}

\title{Reasons, Persons, and Physics}
\author{Luiz Carlos Ryff \\
\textit{Instituto de F\'{\i}sica, Universidade Federal do Rio de Janeiro,}\\
\textit{Caixa Postal 68528, 21041-972 Rio de Janeiro, Brazil}\\
E-mail: ryff@if.ufrj.br}
\maketitle

\begin{abstract}
Starting from general considerations, some ideas of the philosopher Derek
Parfit on consciousness, self-awareness, and reductionism are briefly
reviewed and critically\ examined from the standpoint of physics.
\end{abstract}

\textbf{1. Introduction }\bigskip

The philosopher Derek Parfit, in his book entitled Reasons and Persons 
\textrm{[1]} analyses the difficult subject of consciousness and
self-awareness, introducing the interesting idea of replicas of human
beings. More recently the reductionist approach, which aims at explaining
mind states in terms of the brain functioning has been seriously discussed 
\textrm{[2]}, and quantum approaches to the subject have also been proposed 
\textrm{[3]}. In this essay, using simple and general arguments based on
self-awareness (namely the subjective\ feeling of existing or being at a
certain time and in a certain place), I wish to discuss\ the point of view
according to which consciousness is a consequence of physical processes. As
we will see, this raises difficult questions that seem to indicate that
contemporary physics is unable to solve them.

\bigskip\ 

\textbf{2. The Continuity of the Self}

\bigskip

Let us, initially, briefly discuss some aspects we believe are inherent to
consciousness and try to clarify what is to be understood by reduction of
consciousness to physical processes. Even if we finally obtain a physical
description of consciousness, do we really know what consciousness is?
Naturally, to really know what consciousness is we have to be conscious or
aware. No physical explanation can be a substitute for this personal
experience, to which philosophers refer to as the hard problem. In this
respect, no explanation of consciousness can ever exist. However, we can
assume that consciousness is a consequence of the physical state of a
person, and conjecture that if it were possible to have two identical and
indistinguishable individuals they would have to be, or at least to feel as
if they were, the very same person (we will return to this subject below).
Naturally, this immediately raises a question: If these people are in
different space regions, they have different histories and are under
different physical conditions; from this point of view, they cannot,
strictly speaking, be identical individuals. However, we always assume a
sort of continuity of the self. For example, if a person can either stay at
home or go to the theatre, he or she will not become another individual
because he or she decided to go, or not to go. There are many different
physical situations or states which do not change our feeling of being
ourselves. We simply know that we are in a different place or under
different circumstances, but we are still the very same person. Therefore, I
will assume that there can, in principle, be two physically identical
individuals, in the sense of being indistinguishable from each other, even
though they can be in different regions in space.

\bigskip

\textbf{3. Is it Possible, in Principle, to Duplicate Consciousness?}

\bigskip

The following gedanken experiment, in which we assume the continuity of the
self, will make it evident that the reductionist approach raises some
interesting and difficult conceptual questions. If consciousness is the
result of physical processes, it is possible, in principle, to have two 
\textit{different} individuals (\textit{different} in the sense of being in
two different and distant places, for example) with the same
self-consciousness. We can try to make the argument more dramatic by
introducing the following \textit{Duplicating Machine}. Let us imagine that,
in the very distant future, scientists have developed a machine that makes
perfect replicas of individuals. They are interested in sending people to
Mars, to try to turn it into a hospitable place. However, because of the
very hostile conditions in this planet, there are no candidates. To
circumvent this difficulty, they suggest the following scheme. The possible
candidate will enter the duplicating machine, which will generate two
indistinguishable individuals. It is impossible to know which one is the
original and which one is the replica: they are identical to the candidate.
One of them will be sent to Mars, where he/she will not have a very happy
life, while the other, who will stay on Earth, will be recompensed. The
question then is: is it an advantage to enter the duplicating machine? Will
the candidate identify him/herself with the person who stays on Earth or
with the person who is sent to Mars? We are interested in the possible
answer to the following question: do the two individuals share the same
self-consciousness? Actually, what might it be like `to share the same
self-consciousness'? Could the very same person have the sensation of being
at two different places at the same time, as if his/her consciousness were
split? If the answer is \textbf{yes}, we will have problems with special
relativity; and, at the same time, it reminds us of the quantum
entanglement, where two systems can be instantaneously connected without
apparently any kind of physical interaction between them (more about this in
section 5). On the other hand, if the answer is \textbf{no}, there seem to
be the following possibilities: (1) consciousness cannot be reduced to
physical processes, and (2) matter has some sort of `proto-consciousness',
that is, two people can be identical and have different consciousness
because they are made of different material particles. But conclusion (2)
hardly seems to be consistent with the continuity assumption and the fact
that the atoms in our body are continually being changed. A third
possibility (to be discussed in section 5) is to consider the continuity of
the self as a sort of illusion (but who would be being deluded?).

\bigskip

\textbf{4. The Paradox of Reductionism}

\bigskip

Intimately connected with the problem of consciousness is the problem of
free will, since there seems to be no reason to be conscious if there is no
free will. In particular, consciousness would not be an advantage in the
process of natural selection, as it seems to be. Every scientific project
presupposes free will: we have to discuss our plans and assume, on the basis
of the arguments that have been presented, that we are free to decide which
experiment to perform and to establish our scientific policy. However, if
our scientific project includes the reduction of our behavior to physical
processes, we apparently face a contradiction. If our behavior is the
consequence of `blind' physical laws, whether they are deterministic or
probabilistic, it does not matter, there is no room for free will, and, as a
consequence, any scientific project would be meaningless. In other words, to
develop a physical project we have to assume that there must be some
processes which cannot be reduced to physics, that is, they cannot be
included in the project \textrm{[4]}.

\bigskip

\textbf{5. Discussion}

\bigskip

Starting from the assumption of the continuity of the self, we have
examined, using simple and general considerations, some possible
consequences of the idea according to which consciousness results from
physical processes inside the brain. By continuity of the self is to be
understood the consciousness feature that makes a person, at different
places, on different occasions, and in different circumstances, still feel
as being the very same individual. We have discussed the in-principle
possibility of having two physically indistinguishable individuals, and
posed the question of whether or not they would share the same split
self-consciousness. If the answer is affirmative, there seems to be a
problem with special relativity. For example, if one of the two individuals
is destroyed, we can have a frame in which the two still coexist \textrm{[5]}%
. In this case, is self-consciousness still split or not? In which frame? It
seems that a privileged frame has to be introduced \textrm{[6]}. This has a
resemblance with the difficulties we find in quantum mechanics whenever we
try to ascribe an objective reality to the state of a system that is
entangled with another system \textrm{[7]}. In this case two systems can be
instantaneously connected without apparently any kind of physical
interaction between them. Similarly, if the two indistinguishable
individuals share the same self-consciousness somehow they must be
entangled. \ Although this may sound as a farfetched and preposterous idea,
the alternative (there is no split of the self-consciousness), is not
without its own difficulties. In this case, there seems to be the following
possibility: there is no splitting of self-consciousness (or, more
specifically, self-awareness) because the two individuals, although
indistinguishable, are made of different molecules, atoms and particles. In
this case, sheer matter would be endowed with a sort of proto-consciousness.
In some sense, our self-consciousness would be the consciousness of being
made of definite individual particles. However, this idea can hardly be
accommodated with the assumption of the continuity of the self and the fact
that the atoms in our body are continually being changed. Strictly speaking,
it amounts to rejecting the possibility of reducing consciousness to
physical processes. We know that parts of our body (even the heart) can be
replaced, without any modification occurring in our inner feeling of
self-awareness. We still remain the very same person. We can take a step
further and imagine the following fictitious scenario: if we exchange part
of a person's brain for another totally identical part, but composed of
different atoms, would we have, as a final result, a person possessing a
different self-consciousness? We might be tempted to borrow the idea of
decoherence from quantum mechanics \textrm{[8]}. Two initially identical
individuals become, in an extremely short time, two different individuals,
because of decoherence. However, this is also hardly consistent with the
continuity of the self. Moreover, it is far from clear that the decoherence
approach can satisfactorily explain actualization in quantum mechanics 
\textrm{[9]}, that is, the transformation of \textit{and }into \textit{or}.
Strictly speaking, as consequence of entanglement the initial \textit{and}
becomes \textit{and} + \textit{and} + \textit{and} + ... .

Perhaps, a more radical and (certainly) disputable point of view would be to
assume that the continuity of the self is a sort of illusion. `We' are
continually becoming other individuals, with the `illusion' of still being
the very same person. The following example may clarify this point. Let us
imagine a car whose different parts are continually replaced. After some
time, we have a totally new car which looks exactly like the initial one and
to which we refer as being the same car. The very same thing would happens
to us. In some sense, we would be replicas of ourselves. We would be
continually `dying' and being `born'. A person would die when this process
of `rebirth' is interrupted. From this point of view it does not make any
difference for the candidate that enters the duplicating machine whether the
replica will be sent to Mars or not. The replica would not be the candidate
anyway. Actually, there would be two indistinguishable individuals; even so,
there would be some aspect that would make each one essentially different
from the other, since they would have independent self-consciousness. From
this way of thinking\textrm{\ }-- highly\ questionable, in my opinion -- the
continuity of the selves can be seen as being equivalent to a succession of
replicas. In some sense, we would live only a short time, determined by the
time duration of the processes inside the brain responsible for the
self-consciousness feeling. In other words, the plans we make for our future
are actually\ not plans for \textit{our} future, they are plans we make for
replicas that will live in the future and will inherit our memories.

We have also seen that the reductionist approach leads to a sort of paradox.
To the extent that it presupposes a scientific project, it assumes free
will. But blind laws of physics, ruled by mathematical equations, with no
purpose, leave no room for free will. Actually, we have no choice to decide
about our genetic inheritance and about the place, the time, and the
circumstances in which we are born, namely factors that shape our behavior.
So maybe we are not free after all. But then, what is the reason for being
conscious, and feeling pain and pleasure, and being sad and happy?

Very probably, in order to unify mental and physical processes, we will need
a physics in which, somehow, features implicitly connected with mental
processes are already incorporated in its foundations, which is not exactly
the same as saying that mental processes will be reduced to physics. Making
physical copies, or replicas, of minds may be an impossibility, even in
principle. Each individual is connected to the environment. As a
consequence, although it may be possible, at least in principle, to
substitute the human mind by an identical physical structure made of
different atoms, it may not be possible to exactly remake this external
connection that results of the personal history of each individual. In this
respect, a human being would not be capable of being reduced to its physical
components. When you perform this replacement you break the external
connections.

It is interesting to highlight the active role of the scientist in the use
of the laws of physics, which can make room for the introduction of mental
processes into these very laws. The laws of physics can be interpreted as
establishing a program, which consists of seeking to fit the facts into a
given structure. In the case of classical physics, and taking Newton's
second law of dynamics as an example, we assume that, given a concrete
situation, it is always possible, at least in principle, to find an\textrm{\ 
}$F$ and an\textrm{\ }$m$\textrm{\ }that satisfy the equation\textrm{\ }$F=ma%
\mathrm{.}$ We don't need to get involved in a vicious circle trying to
define\textrm{\ }$F$\textrm{\ }in terms of\textrm{\ }$m$\textrm{\ }and%
\textrm{\ }$m$\textrm{\ }in terms of\textrm{\ }$F$\textrm{.\ }Similarly, in
quantum mechanics we assume that it is always possible to associate a state
vector with a system, that this vector will evolve satisfying the Schr\"{o}%
dinger equation, and that, given a concrete situation, we can identify a
projective measurement. This explains why quantum mechanics is so effective,
despite its widely varying interpretations.


\begin{thebibliography}{1}
\bibitem[4]{} The collapse of the state vector in quantum mechanics might be
an indication that the reductionist approach is inevitably limited.
\bibitem[5]{} We know that two events separated by a space-like interval
cannot be causally connected, and that absolute simultaneity can only exist
for events that occur at the same time at the same place. We can imagine the
two indistinguishable\ individuals distant from each other and in different
Lorentz frames.
\end{thebibliography}
\end{document}